\documentclass[10pt,a4paper]{article}
\usepackage{times}
\usepackage{a4wide}
\usepackage{amsfonts}
\usepackage{amssymb}
\usepackage{amsmath}
\usepackage{ifpdf}
\ifpdf
\usepackage[pdftex,unicode,implicit]{hyperref}
\usepackage[utf8]{inputenc}
\usepackage[english]{babel}
\usepackage{amsmath}
\usepackage{amsfonts}
\usepackage{amssymb}
\usepackage{epsfig}
\usepackage{graphics,psfrag,rotating}
\usepackage{graphicx}
\usepackage{dcolumn}
\usepackage{bm}
\usepackage{epstopdf}
\usepackage{epsfig}
\usepackage{color}
\usepackage[usenames,dvipsnames,svgnames]{xcolor}

\usepackage{concmath}
\usepackage[T1]{fontenc}

\DeclareSymbolFont{extraup}{U}{zavm}{m}{n}
\DeclareMathSymbol{\varheart}{\mathalpha}{extraup}{86}
\DeclareMathSymbol{\vardiamond}{\mathalpha}{extraup}{87}

\usepackage[sort&compress,comma]{natbib}
\bibpunct{[}{]}{,}{n}{}{,}

\usepackage[a4paper,bindingoffset=0.2in,left=0.6in,right=0.6in,top=1in,bottom=1in]{geometry}

\hypersetup{%
  pdftitle    = {Topological black holes in ungauged Supergravity},
  pdfkeywords = {topological black holes, FGK formalism, supergravity, String Theory},
  pdfauthor   = {C.S. Shahbazi},
  plainpages  = true,
  colorlinks  = true,
  citecolor   = blue,
  urlcolor    = red,
  linkcolor   = black
}

\else
  \usepackage[dvips]{graphicx}
  \usepackage[unicode,implicit]{hyperref}

\fi
\makeatletter
\@addtoreset{equation}{section}
\makeatother

\begin{document}


\vspace*{-1.3cm}
\newcommand{\HRule}{\rule{\linewidth}{0.3mm}}
\setlength{\parindent}{1cm}
\setlength{\parskip}{1mm}
\noindent
\HRule
\begin{center}

\vspace{1cm}

{\LARGE {\bf Topological solutions in ungauged Supergravity}}\\[8mm]
\HRule
\vspace{1.5cm}

\begin{center}

\renewcommand{\thefootnote}{\alph{footnote}}


{A.\,de la Cruz-Dombriz$^{\vardiamond,\clubsuit}  \footnote{E-mail: {\tt  alvaro.delacruzdombriz [at] uct.ac.za}}$, M.~Montero$^{\varheart} \footnote{E-mail: {\tt mig.montero [at] estudiante.uam.es}}$ 
and C. S.~Shahbazi$^{\varheart} \footnote{E-mail: {\tt Carlos.Shabazi [at] uam.es}}$
}
\\
\renewcommand{\thefootnote}{\arabic{footnote}}

\vspace{.5cm}

{\it
$^{\vardiamond}$  Astrophysics, Cosmology and Gravity Centre (ACGC) and
Department of Mathematics and Applied Mathematics, University of Cape Town, Rondebosch 7701, Cape Town, South Africa.
\\
$^{\clubsuit}$ Instituto de Ciencias del Espacio (ICE/CSIC) and Institut d'Estudis Espacials de Catalunya (IEEC), Campus UAB, Facultat de Ci\`{e}ncies, Torre C5-Par-2a, 08193 Bellaterra (Barcelona) Spain.
\\
$^{\varheart}$ Instituto de F\'{\i}sica Te\'orica UAM/CSIC\\
C/ Nicol\'as Cabrera, 13--15,  C.U.~Cantoblanco, 28049 Madrid, Spain.
%
}


\vspace{.5cm}

\end{center}

{\bf Abstract}

\begin{quotation}

  {\small 
A new general class of solutions of ungauged four-dimensional Supegravity, in one-to-one correspondence with spherically symmetric, static black-hole solutions and Lifshitz solutions with Hyperscaling violation (\emph{hvLif}) is studied. The causal structure of the space-time is then elucidated. 
}

\end{quotation}

\end{center}

\setcounter{footnote}{0}

\newpage
\pagestyle{plain}

\tableofcontents

\vspace{1cm}

\section*{General remarks}

The dimensional reduction performed in \cite{Ferrara:1997tw} allows, when considering a spherically symmetric and static background, to write down the equations of motion of any four-dimensional ungauged Supergravity as an effective, one-dimensional, system of differential equations for the scalars  fields $\left\{\phi^{i}\, , i=1,\dots,n_{v}\right\}$ and the metric warp factor $U$, since the vector fields and one of the two arbitrary functions of the metric can be explicitly integrated.

In \cite{Bueno:2012sd}, a remarkable fact was found: given a solution $(U,\phi^{i})$ of the one-dimensional equations of motion, a solution of the complete four-dimensional theory can be constructed not only using the spherically symmetric, static space-time metric, but also using two other different space-time metrics. In other words, given a solution $(U,\phi^{i})$ of the one-dimensional equations of motion, we can choose three different space-time metrics such that the complete four-dimensional solution obeys the equations of motion of the original theory.     

One of these three choices is, of course, the spherically symmetric and static space-time metric describing a black-hole solution, which we shall denote by $C_1$. The second one was previously investigated in \cite{Bueno:2012sd} and corresponds to Lifshitz solutions with Hyperscaling violation (\emph{hvLif}), and will be denoted by $C_2$. The third and final choice, $C_3$, remains to be completely identified, and its study is the leit-motiv of this note. We shall find that this class of solutions corresponds to a specific kind of naked singularities in either static or time-dependent solutions, depending on the values of the solution's parameters, that we shall illustrate by studying two simple examples. Therefore, the solutions belonging to $C_3$ are not topological black holes, in the sense that it is commonly understood in the literature \cite{Mann:1997iz}. However, they are still topological solutions, i.e., they represent static space-times with a \emph{topological} space-like slicing. In other words, the spacetime is foliated by a family of two-dimensional surfaces, each being locally isometric to the hyperbolic plane, which can in principle be of arbitrary genus, depending on the existence of global identifications as shown in \cite{Mann:1997iz}.

In any case, a \emph{triality} among three general classes ($C_1$, $C_2$ and $C_3$) of solutions in four-dimensional Supergravity can be established in terms of a 1-1-1 map: i.e., for any solution $s_1\in C_1$ there is one and only one corresponding solution $s_2\in C_2/\mathbb{Z}_2$\footnote{The $\mathbb{Z}_2$ identification is needed to relate pairs of solutions in $C_2$ whose transverse part is related by a change of sign in the radial coordinate.} and one and only one solution $s_3\in C_3$ such that $s_1$, $s_2$ and $s_3$ are constructed in terms of the same $(U,\phi^{i})$ appearing in the one-dimensional equations of motion.   

Finally, as a consequence of the triality, all the methods developed to obtain black-hole solutions in ungauged four-dimensional Supergravity \cite{Meessen:2011aa,Meessen:2012su,Galli:2012pt,Mohaupt:2011aa}, as well as the new results concerning the effective one-dimensional equations of motion \cite{Galli:2012ji,Galli:2012jh,Bueno:2013pja}, can be applied to solutions belonging to the classes $C_2$ and $C_3$.

The new class $C_{3}$ of solutions are relevant for several reasons. It is a class of solutions which can be easily embedded in String Theory, for example by means of Type-II flux-less Calabi-Yau compactifications, and therefore they correspond to states in the full-fledged String Theory, after being appropriately corrected. In addition, they are a non-trivial example which exhibits the attractor mechanism, different from all the previous solutions where the attractor mechanism was proven to hold \cite{Ferrara:1996dd,Ferrara:1997tw,Bellorin:2006xr,Bellucci:2007ds,Ferrara:2007qx,Bellucci:2008jq}. The attractor mechanism was of outermost importance in Supergravity and String Theory in order to check the macroscopic computation, at strong coupling, of the entropy of a black hole versus the microscopic calculation, at weak coupling, where the black hole becomes a configuration of D-branes and other objects \cite{Strominger:1996sh,Maldacena:1996gb}. Since it is possible to associate to each black hole solution a unique topological solution, it would be really interesting to see what is the microscopic pricture of these solutions in String Theory. In doing so we could compare the microscopic description of the black hole and the microscopic description of the corresponding topological solution, which will give us information about what corresponds intrinsically to the microscopic picture of a black hole, which possesses an event horizon. Furthermore, the topological solutions provide evidence about the existence of different new brane solutions in higher dimensions, which will provide us with the higher dimensional Supergravity objects necessary to obtain the topological four-dimensional solutions by appropriate intersection and dimensional reduction, very much in the style of what happens for black hole solutions in four dimensions that can be obtained from a particular intersection of brane solutions to Supergravity in higher dimensions. 

This communication is organised as follows: in Section  \ref{sec:FGKgeneralized}
we introduce the Ferrara-Gibbons-Kallosh (F.G.K.) formalism, developed in \cite{Ferrara:1997tw}, and the effective one-dimensional equations of motion governing the theory. 
Section \ref{sec:TopologicalSchwarzschild} is focused in the topological Schwarzschild-like solution\footnote{By ``topological Schwarzschild-like solution'' we mean the solution in $C_{3}$ obtained by using the $(U,\phi^{i})$ effective solution corresponding to the Schwarzschild black hole in $C_{1}$. Similar considerations apply to the topological Reissner-Nordstr\"om-like solution.}, where we distinguish two cases 
depending on the sign of the available arbitrary coefficient. Then in Section \ref{sec:TopologicalRN} we study the topological Reissner-Norstr\"om-like solution and depict its Carter-Penrose diagram. 


\section{The generalised F.G.K. formalism}
\label{sec:FGKgeneralized}


Following Ref.~\cite{Ferrara:1997tw}, let us consider the action
\begin{eqnarray}
\label{eq:generalaction}
I
\!=\!
 \int\!\! {\rm d}^{4}x \sqrt{|g|}
\left(
R +\mathcal{G}_{ij}(\phi)\partial_{\mu}\phi^{i}\partial^{\mu}\phi^{j}+  
 2 \Im{\rm m}\,\mathcal{N}_{\Lambda\Sigma}
F^{\Lambda}{}_{\mu\nu}F^{\Sigma\, \mu\nu}
-2 \Re{\rm e}\,\mathcal{N}_{\Lambda\Sigma}
F^{\Lambda}{}_{\mu\nu}\star F^{\Sigma\, \mu\nu}
\right)\,  ,   
\end{eqnarray}
where $\mathcal{N}_{\Lambda\Sigma}$ is the complex, scalar-dependent,
(\textit{period}) matrix. The bosonic sector of any ungauged supergravity
theory in 4 dimensions can be expressed through this action. The scalars are labeled
by $i,j,\dots~= 1,\dots,n_{s}$, and the vector fields by $\Lambda, \Sigma,\dots~= 0,\dots,n_{v}$. The
scalar metric $\mathcal{G}_{ij}$ and the period matrix
$\mathcal{N}_{\Lambda\Sigma}$ depend on the particular theory under
consideration.

Since we are interested in obtaining static solutions, let us consider the metric

\begin{equation}
\label{eq:generalbhmetric}
{\rm d}s^{2} 
 = 
e^{2U} {\rm d}t^{2} - e^{-2U} \gamma_{\underline{m}\underline{n}}
{\rm d}x^{\underline{m}}{\rm d}x^{\underline{n}}\, , 
\end{equation}

\noindent
where $\gamma_{\underline{m}\underline{n}}$ is a 3-dimensional
(\textit{transverse}) Riemannian metric, to be specified later. Using
Eq.~(\ref{eq:generalbhmetric}) and the assumption of staticity for all the
fields, we perform a dimensional reduction over time in the equations of
motion that follow from the aforementioned general action. Thus, we obtain 
a set of reduced equations of motion that we can write in the form 
\cite{Ferrara:1997tw}
\begin{eqnarray}
\label{eq:Eq3dim1}
\nabla_{\underline{m}}
\left(\mathcal{G}_{AB} \partial^{\underline{m}}\tilde{\phi}^{B}\right)
-\tfrac{1}{2}\partial_{A} \mathcal{G}_{BC}
\partial_{\underline{m}}\tilde{\phi}^{B}\partial^{\underline{m}}\tilde{\phi}^{C} 
& = & 
0\, ,
\\
& & \nonumber \\
\label{eq:Eq3dim2}
R_{\underline{m}\underline{n}}
+\mathcal{G}_{AB}\partial_{\underline{m}}\tilde{\phi}^{A}
\partial_{\underline{n}}\tilde{\phi}^{B} 
& = & 
0\, ,
\\
& & \nonumber \\
\label{eq:Eq3dim3}
\partial_{[\underline{m}}\psi^{\Lambda}\partial_{\underline{n}]}\chi_{\Lambda} 
& = & 
0\, ,
\end{eqnarray}
where all the tensor quantities refer to the 3-dimensional metric
$\gamma_{\underline{m}\underline{n}}$ and where we have defined the metric
$\mathcal{G}_{AB}$ as follows
\begin{equation}
\mathcal{G}_{AB}
\equiv
\left(
  \begin{array}{ccc}
   2 &  &  \\
  & \mathcal{G}_{ij} &  \\
  &   & 4 e^{-2U}\mathcal{M}_{MN} 
  \end{array}
\right)\, ,
\end{equation}
in the \emph{extended} manifold of coordinates
$\tilde{\phi}^{A}=\left(U,\phi^{i},\psi^{\Lambda},\chi_{\Lambda}\right)$,
where
\begin{equation}
(\mathcal{M}_{MN})
\equiv
\left( 
\begin{array}{lr}
    (\mathfrak{I}+\mathfrak{R}\mathfrak{I}^{-1}\mathfrak{R})_{\Lambda\Sigma} &
    -(\mathfrak{R}\mathfrak{I}^{-1})_{\Lambda}{}^{\Sigma} \\
    & \\
    -(\mathfrak{I}^{-1}\mathfrak{R})^{\Lambda}{}_{\Sigma} &
    (\mathfrak{I}^{-1})^{\Lambda\Sigma} \\   
  \end{array}
\right)
\, ,
\hspace{.3cm}
\mathfrak{R}_{\Lambda\Sigma} \equiv \Re{\rm e}\,\mathcal{N}_{\Lambda\Sigma}\; {\rm and}
\hspace{.3cm}
\mathfrak{I}_{\Lambda\Sigma} \equiv \Im{\rm m}\,\mathcal{N}_{\Lambda\Sigma}\, .
\end{equation}
Eqs.~(\ref{eq:Eq3dim1}) and (\ref{eq:Eq3dim2}) can be obtained from the
three-dimensional effective action
\begin{equation}
\label{eq:Eq3dim3action3dim}
I=\int {\rm d}^{3}x \sqrt{|\gamma|}
\left\{ R 
+\mathcal{G}_{AB}\partial_{\underline{m}}\tilde{\phi}^{A}
\partial^{\underline{m}}\tilde{\phi}^{B}\right\}\, ,
\end{equation}
once the constraint given by 
Eq.~(\ref{eq:Eq3dim3}) has been added.

In order to further dimensionally reduce the theory to a mechanical one-dimensional problem, we introduce the following transverse metric 
\begin{equation}
\label{eq:gammak}
\gamma_{\underline{m}\underline{n}}
{\rm d}x^{\underline{m}}{\rm d}x^{\underline{n}}
 = 
\frac{{\rm d}\tau^{2}}{W_{\kappa}^{4}} 
+
\frac{{\rm d}\Omega^{2}_{\kappa}}{W^{2}_{\kappa}}\, ,
\end{equation}
where $W_{\kappa}$ is a function of $\tau$ and ${\rm d}\Omega^{2}_{\kappa}$ is the metric of
the 2-dimensional symmetric space of curvature $\kappa=-1,0,1$ and unit radius respectively as follows
\begin{eqnarray}
{\rm d}\Omega^{2}_{(1)} & \equiv & {\rm d}\theta^{2}+\sin^{2}{\! \theta}\, {\rm d}\phi^{2}\, ,
\\
& & \nonumber \\
\label{eq:domega1}
{\rm d}\Omega^{2}_{(-1)} & \equiv & {\rm d}\theta^{2}+\sinh^{2}{\! \theta}\, {\rm d}\phi^{2}\, ,
\\
& & \nonumber \\
\label{eq:domega0}
{\rm d}\Omega^{2}_{(0)} & \equiv & {\rm d}\theta^{2}+{\rm d}\phi^{2}\, .
\end{eqnarray}
In these three cases the $(\theta,\theta)$ or the $(\phi,\phi)$ component of the Einstein equations can be solved for $W_{\kappa}(\tau)$, giving
\begin{eqnarray}
\label{eq:sinh}
W_{1} 
& = & 
\frac{\sinh{r_{0} \tau}}{r_{0}}\, ,\\
& & \nonumber \\
\label{eq:cosh}
W_{-1}
& = &
\frac{\cosh{r_{0} \tau}}{r_{0}}\, , \\
& & \nonumber \\
\label{eq:exp}
W^{\pm}_{0}
& = & 
a e^{\mp r_{0}\tau}\, .
\end{eqnarray}
where $a$ is an arbitrary real constant with dimensions of inverse length and $r_0$ is an integration constant whose interpretation depends on $\kappa$. The case $\kappa=1$ 
has been widely studied in the literature and corresponds to asymptotically flat, spherically symmetric, static black holes \cite{Ferrara:1997tw,Galli:2011fq,Mohaupt:2011aa,Ferrara:2008hwa}. The case $\kappa=0$ has been recently studied in \cite{Bueno:2012sd} and provides a rich spectrum corresponding to Lifshitz-like solutions with hyper-scaling violation. Thus, the goal of this letter is to study the case $\kappa=-1$. 

For the three cases (\ref{eq:sinh}), (\ref{eq:cosh}) and (\ref{eq:exp}) we are left with the same  equations for the one-dimensional fields, which can be written as follows
\begin{eqnarray}
\label{eq:Eq1tau}
\frac{{\rm d}}{{\rm d}\tau} \left(\mathcal{G}_{AB}
  \frac{{\rm d}\tilde{\phi}^{B}}{{\rm d}\tau}\right)
-\tfrac{1}{2}\partial_{A}\mathcal{G}_{BC}\frac{{\rm d}\tilde{\phi}^{B}}{{\rm d}\tau}
\frac{{\rm d}\tilde{\phi}^{C}}{{\rm d}\tau}
& = & 
0\, ,
\\
& & \nonumber \\
\label{eq:Eq2tau}
\mathcal{G}_{BC}\frac{{\rm d}\tilde{\phi}^{B}}{{\rm d}\tau}
\frac{{\rm d}\tilde{\phi}^{C}}{{\rm d}\tau}
& = & 
2 r^{2}_{0}\, .
\end{eqnarray}
The electrostatic and magnetostatic potentials $\psi^{\Lambda},\chi_{\Lambda}$
only appear through their $\tau$-derivatives. The associated conserved
quantities are the magnetic and electric charges
$p^{\Lambda},q_{\Lambda}$ that can be used to eliminate completely the
potentials. The remaining equations of motion can be reorganized in the convenient
form 
\begin{eqnarray}
\label{eq:e1}
U^{\prime\prime}
+e^{2U}V_{\rm bh}
& = & 0\, ,\\ 
& & \nonumber \\
\label{eq:Vbh-r0-real}
(U^{\prime})^{2} 
+\tfrac{1}{2}\mathcal{G}_{ij}\phi^{i\, \prime}  \phi^{j\, \prime}  
+e^{2U} V_{\rm bh}
& = & r_{0}^{2}\, ,\\
& & \nonumber \\
\label{eq:e3}
(\mathcal{G}_{ij}\phi^{j\, \prime})^{\prime}
-\tfrac{1}{2} \partial_{i}\mathcal{G}_{jk}\phi^{j\, \prime}\phi^{k\, \prime}
+e^{2U}\partial_{i}V_{\rm bh}
& = & 0\, ,
\end{eqnarray}
in which the prime indicates differentiation with respect to $\tau$ and the
so-called \textit{black-hole potential} $V_{\rm bh}$ is given by
\begin{equation}
V_{\rm bh}(\phi,\mathcal{Q})
\equiv
\tfrac{1}{2}\mathcal{Q}^{M}\mathcal{Q}^{N} \mathcal{M}_{MN}\, ,
\hspace{1cm}
(\mathcal{Q}^{M})
\equiv
\left(
  \begin{array}{c}
   p^{\Lambda} \\ q_{\Lambda} \\ 
  \end{array}
\right)\, .
\end{equation}
Eqs.~(\ref{eq:e1}) and (\ref{eq:e3}) can be in fact derived from the effective action
\begin{equation}
\label{eq:effectiveaction}
I_{\rm eff}[U,\phi^{i}] = \int {\rm d}\tau \left\{ 
(U^{\prime})^{2}  
+\tfrac{1}{2}\mathcal{G}_{ij}\phi^{i\, \prime}  \phi^{j\, \prime}  
-e^{2U} V_{\rm bh}
  \right\}\, ,  
\end{equation}
whereas Eq.~(\ref{eq:Vbh-r0-real}) is nothing but the conservation of the Hamiltonian
(due to the absence of explicit $\tau$-dependence in the Lagrangian) with a
particular value of the integration constant $r_{0}^{2}$.

A large number of solutions of the system (\ref{eq:e1}), (\ref{eq:Vbh-r0-real}) and (\ref{eq:e3}), for different theories of $\mathcal{N}=2,d=4$ supergravity coupled to vector supermultiplets, have been found (see \textit{e.g.}~Ref.~\cite{Galli:2011fq,Galli:2012pt,Mohaupt:2011aa,Meessen:2011aa,Breitenlohner:1987dg,
Bergshoeff:2008be,Bossard:2009at,Chemissany:2009hq,Chemissany:2010ay,
Chemissany:2010zp,Chemissany:2012nb,Bellorin:2006yr,Shmakova:1996nz}), always focusing on the case $\kappa=1$. With this choice of transverse metric, they describe
single, charged, static, spherically-symmetric, asymptotically-flat and non-extremal black holes. These solutions can now be studied setting $\kappa=-1$ in the transverse metric. 

Using Eqs. (\ref{eq:domega1}) and (\ref{eq:cosh}), the metric can be written in this case as 
\begin{equation}
\label{eq:generalbhmetriccosh}
{\rm d}s^{2}
= 
e^{2U} {\rm d}t^{2} - e^{-2U}\left[ \frac{r^4_0 {\rm d}\tau^2}{\cosh^4 r_0\tau} 
+
 \frac{r^2_0}{\cosh^2 r_0\tau}{\rm d}\Omega^{2}_{(-1)} \right] \, ,  
\end{equation}

\noindent
where ${\rm d}\Omega^{2}_{(-1)}  = {\rm d}\theta^{2}+\sinh^2\theta\, {\rm d}\phi^{2}$ is the two-dimensional metric of negative constant curvature.


\section{The topological Schwarzschild black hole}
\label{sec:TopologicalSchwarzschild}


The formalism developed in Section \ref{sec:FGKgeneralized} applies to any Lagrangian of the form (\ref{eq:generalaction}). In particular, it can be applied to the case where there are no matter-fields and only the Hilbert-Einstein term remains. In this case, we are dealing with the Einstein equations in vacuum, and we obtain \cite{Galli:2011fq}

\begin{equation}
\label{eqschwarzschildU}
U=\sigma\tau\, ,  
\end{equation}

\noindent
where $\sigma$ is an arbitrary integration constant, which is equal to the mass of the black hole in the asymptotically flat, spherically symmetric, static case. Thus the metric is given by

\begin{equation}
\label{eq:Schwarzschildtau}
{\rm d}s^{2}
= 
e^{2\sigma \tau} {\rm d}t^{2} - e^{-2\sigma \tau}\left[ \frac{\sigma^4 {\rm d}\tau^2}{\cosh^4 \sigma\tau} 
+
 \frac{\sigma^2}{\cosh^2 \sigma\tau}\left({\rm d}\theta^{2}+\sinh^2\theta {\rm d}\phi^{2} \right)  \right] \, .  
\end{equation}

\noindent
In order to write Eq. (\ref{eq:Schwarzschildtau}) in a more convenient way we performe the following change of variables

\begin{equation}
\label{eq:Schwarzschildchange}
e^{2\sigma\tau} = \frac{2 \sigma}{r}-1 \, .  
\end{equation}

\noindent
In these new coordinates, the metric reads

\begin{equation}
\label{eq:Schwarzschildr}
{\rm d}s^{2}
= 
\left(\frac{2\sigma}{r}-1\right){\rm d}t^{2} -\left(\frac{2\sigma}{r}-1\right)^{-1}{\rm d}r^2 - r^2 {\rm d}\Omega^{2}_{(-1)} \, .  
\end{equation}

\noindent
Thanks to Eq. (\ref{eq:Schwarzschildr}) it is easy to recognise the last metric as the so-called $AII$ metric with $\kappa=-1$, found in \cite{EK} and whose interpretation was first given in \cite{Gott:1974yc, Louko:1987gg}. We summarise now the principal properties of such space-time, closely following \cite{GP}, where a detailed description is given.

\subsection{Carter-Penrose diagram}
 
Since $r=0$ is a true singularity, it is convenient to take $r\in \left(0,\infty\right)$, allowing $\sigma$ to be either positive or negative. We 
have therefore two different possibilities, that shall be considered separately.

\begin{enumerate}

\item $\sigma>0$

The metric  (\ref{eq:Schwarzschildr}) can be written as follows

\begin{equation}
\label{eq:Schwarzschildrn}
{\rm d}s^{2}
= 
\left(\frac{2|\sigma|}{r}-1\right){\rm d}t^{2} -\left(\frac{2|\sigma|}{r}-1\right)^{-1}{\rm d}r^2 - r^2 {\rm d}\Omega^{2}_{(-1)} \, .  
\end{equation}

\noindent
For $r>2\sigma$ the metric is time-dependent, since the $r$ coordinate becomes time-like. In $r = 2\sigma$ we have Killing horizon related to $\partial_{t}$. The metric is static for $0<r<2\sigma$. The corresponding Penrose diagram is shown in Fig. \ref{Figure_Penrose_diagram}, taking $\sigma_1=\vert \sigma\vert$ and $\sigma_2=0$ in \eqref{eq:RNr2} in order to recover \eqref{eq:Schwarzschildrn}. It is similar to the Penrose diagram of the Schwarzschild solution \cite{EK} except for a quarter-turn tilting. This is explained by the fact that \eqref{eq:Schwarzschildrn} at constant $\phi,\theta$ is related to the Schwarzschild metric with the same restriction by an overall sign. This reverses the notions of space and time-like vectors from one metric to the other, leaving everything else unchanged. The tilt is explained then by the fact that in Penrose diagrams time-like directions are represented upwards.

$r=0$ represents two different time-like, naked singularities, as is apparent from Fig. \ref{Figure_Penrose_diagram}: The coordinate singularity at $r=2\sigma$ is not an usual event horizon (although as stated above, it is a Killing horizon), since events inside it can be seen from observers near the asymptotic future. In contrast, events taking place in this region cannot be seen from the inside, although events taking place near the asymptotic past can be seen from $r<2\sigma$. It is possible for a particle to travel from past null infinity to future null infinity without ever encountering a singularity. Notice that $r=2\sigma$ is still a Killing horizon.

\item $\sigma<0$

We can write the metric (\ref{eq:Schwarzschildr}) as follows

\begin{equation}
\label{eq:Schwarzschildt}
{\rm d}s^{2}
= 
\left(-\frac{2|\sigma|}{r}-1\right){\rm d}t^{2} -\left(-\frac{2|\sigma|}{r}-1\right)^{-1}{\rm d}r^2 - r^2 {\rm d}\Omega^{2}_{(-1)} \, ,  
\end{equation}

\noindent
and we immediately see that there is no coordinate horizon at $r=2\sigma$, the coordinates behave properly all the way to the singularity. Also, $\partial_{r}$ is now everywhere time-like. Relabeling the coordinates accordingly we obtain

\begin{equation}
\label{eq:Schwarzschildt2}
{\rm d}s^{2}
= 
\left(1+\frac{2|\sigma|}{t}\right)^{-1}{\rm d}t^2 -\left(1+\frac{2|\sigma|}{t}\right){\rm d}r^{2} - t^2 {\rm d}\Omega^{2}_{(-1)} \, .  
\end{equation}

\noindent
The physical singularity is, therefore, at $t=0$. In this case, the corresponding Penrose diagram can be seen in \cite{GP}. The solution may be regarded as a vacuum spatially homogeneous but anisotropic cosmological model that is of Bianchi type III, in which $r$ is a global time coordinate.
 
 \end{enumerate}
 

\section{The topological Reissner-Nordstr\"om black hole}

\label{sec:TopologicalRN}


The Reissner-Nordstr\"om black hole  can be embedded in pure $\mathcal{N}=2,d=4$ supergravity. The metric function of this solution in the $\tau$
coordinates is \cite{Galli:2011fq}
\begin{equation}
e^{-2U} 
= 
\frac{\left(M \cosh r_0 \tau - r_0 \sinh r_0 \tau\right)^2}{r_0^2}\, ,\qquad r^{2}_{0} = M^2-V_{\rm bh}\, .
\end{equation}

\noindent
As in the previous case, we perform a change coordinates 

\begin{equation}
\label{eq:RNchange}
r = - r_0 \tanh r_0 \tau + M \, ,  
\end{equation}
in order to rewrite the metric in a more convenient form. Thus
the metric is given by
\begin{equation}
\label{eq:RNr}
{\rm d}s^{2}
= 
\left(-1+\frac{2 M}{r}-\frac{V_{\rm bh}}{r^2}\right) {\rm d}t^{2} -\left(-1+\frac{2 M}{r}-\frac{V_{\rm bh}}{r^2}\right)^{-1} {\rm d}r^2 -r^2 {\rm d}\Omega^{2}_{(-1)} \, ,  
\end{equation}

\noindent
where 
\begin{equation}
V_{{\rm bh}}=-q^2-\frac{p^2}{4}\label{Vbh}\, ,  
\end{equation}
is the black-hole potential of pure $\mathcal{N}=2,d=4$ Supergravity in the chosen
conventions \cite{Galli:2011fq}. The parameters $M$ and $V_{{\rm bh}}$ have a clear physical interpretation in the spherically symmetric case, which however may not carry over to the $\kappa = -1$ case. Therefore we rewrite Eq. (\ref{eq:RNr}) as

\begin{equation}
\label{eq:RNr2}
{\rm d}s^{2}
= 
\left(-1+\frac{2 \sigma_1}{r}+\frac{\sigma^2_2}{r^2}\right) {\rm d}t^{2} -\left(-1+\frac{2 \sigma_1}{r}+\frac{\sigma^2_2}{r^2}\right)^{-1} {\rm d}r^2 -r^2 {\rm d}\Omega^{2}_{(-1)} \, ,  
\end{equation}

\noindent
where $\sigma_1$ and $\sigma_2$ are arbitrary real parameters. Remarkably, the causal structure of the spacetime is independent of the particular values of $\sigma_1,\sigma_2$.

\subsection{Carter-Penrose diagram}

The causal structure of more general cases, in the presence of non-trivial scalars, is analogous to the Topological Reissner-Nordstr\"om-like solution, which is therefore the relevant example  which allows us to identify the space-time features of the whole class of solutions, exactly as in the spherically symmetric case.

\noindent
This spacetime exhibits a physical singularity at $r=0$. Therefore it is enough to restrict ourselves to $r>0$, while allowing $\sigma_1$ to take any value. For the study of the Carter-Penrose diagram, let us remember that the metric \eqref{eq:RNr2} possesses two  Killing horizons,

\begin{equation}
r_{\pm}\equiv \sigma_1 \pm \sqrt{\sigma_1^2+\sigma_2^2}
\end{equation}

\noindent
and only one of these,  

\begin{equation}
r_{+}=r_H\equiv 
	\begin{cases}
		\sigma_1 + \sqrt{\sigma_1^2+\sigma_2^2}  & \mbox{if } \sigma_1 \geq 0 \\
		-\sigma_1 + \sqrt{\sigma_1^2+\sigma_2^2} & \mbox{if } \sigma_1 < 0
	\end{cases}
\end{equation}

\noindent
is greater than zero. That means that there is only one Killing horizon associated to $\partial_t$. For $r>r_H$ the metric is time-dependent, whereas for $0<r<r_H$ it is static. This is the same behaviour of the type AII metric \eqref{eq:Schwarzschildrn} for $\sigma > 0$. Indeed, on their respective $\theta,\phi$ constant slices, these two spacetimes are related by a conformal transformation and hence have the same causal structure and Carter-Penrose diagram. To see this, notice that the metric \eqref{eq:RNr2} is related by a global sign to the Reissner-Nordstr\"om metric with an imaginary value of the charge. Following \cite{GP}, we may introduce Kruskal-Skezeres-like coordinates as follows
\begin{align}
U_+&=-\frac{2r_H^2}{r_H-r_-}\left\vert\frac{r}{r_H}-1\right\vert^{1/2}\left\vert\frac{r}{r_-}-1\right\vert^{\frac{r^2_-}{2r_H^2}}\exp\left[-\frac{(r_H-r_-)}{2r_H^2}(t-r)\right],\nonumber\\ 
V_+&=\frac{2r_H^2}{r_H-r_-}\left\vert\frac{r}{r_H}-1\right\vert^{1/2}\left\vert\frac{r}{r_-}-1\right\vert^{\frac{r^2_-}{2r_H^2}}\exp\left[\frac{(r_H-r_-)}{2r_H^2}(t+r)\right],
\end{align}
in terms of which \eqref{eq:RNr2} takes the form
\begin{align}
{\rm d}s^{2}
&=4\frac{r_-r_H}{r^2}\left\vert\frac{r-r_-}{r_-}\right\vert^{1+\frac{r_-^2}{r_H^2}}\exp\left(-\frac{r_H-r_-}{r_H^2}r\right){\rm d}U_+{\rm d}V_+ -r^2 {\rm d}\Omega^{2}_{(-1)}\nonumber\\
&=\Omega(r;\,r_{H},\,r_{-})\left[-4\frac{r_H}{r}\exp\left(-\frac{r}{r_H}\right){\rm d}U_+{\rm d}V_+\right] -r^2 {\rm d}\Omega^{2}_{(-1)},
\label{3.9}
\end{align}
with
\begin{equation}
\Omega(r;\,r_{H},\,r_{-}) \,\equiv\,\frac{-r_-}{r}\left\vert\frac{r-r_-}{r_-}\right\vert^{1+\frac{r_-^2}{r_H^2}}\exp\left({\frac{r_-}{r_H^2}r}\right).
\end{equation}
The factor multiplied by $\Omega(r;\,r_{H},\,r_{-})$ in the expression (\ref{3.9}) corresponds to the $t-r$ part of the metric \eqref{eq:Schwarzschildrn} in  Kruskal-Skezeres-like coordinates. Since $\Omega(r;\,r_{H},\,r_{-})>0$ is well defined throughout the spacetime, this shows the conformal equivalence between the two metrics in the $\theta,\phi$ constant slices.

\begin{figure*}[htbp]
\label{fig:penrosediagram}
	\centering
			\includegraphics[width=0.6\textwidth]{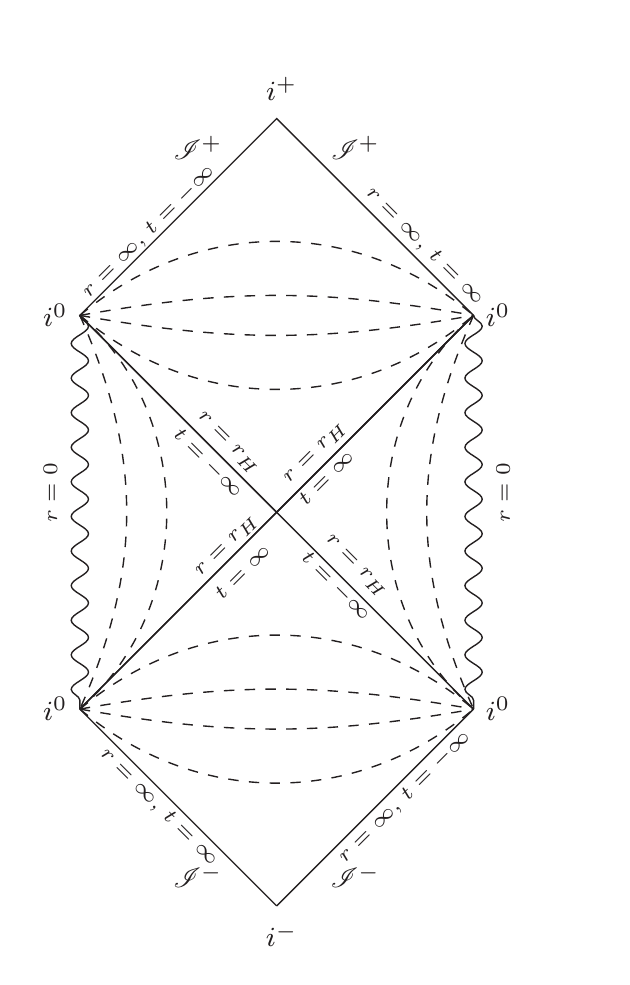}
		
\caption{\footnotesize{Conformal diagram for metric (\ref{eq:RNr2}) with $\sigma_{1}>0$ . This represents sections on which angular coordinates are constant, so that each point on the diagram represents a point on an topological surface of constant negative curvature. The symbols in the figure possess their standard interpretation in this kind of diagrams.}}
\label{Figure_Penrose_diagram}
\end{figure*}

This equivalence of conformal structures can be understood on physical grounds by considering  \eqref{eq:RNr2} with a global sign change. As stated above, this corresponds to a Reissner-Nordstr\"om metric with an imaginary charge. This results in an attractive instead of a repulsive singularity at short distances, which will behave qualitatively in the same way as in Schwarzschild spacetime. Therefore, (\ref{eq:Schwarzschildrn}) and (\ref{eq:RNr2}) share the same Carter-Penrose diagram, given by Figure \ref{Figure_Penrose_diagram}. 

The solution \eqref{eq:RNr2} can be given a physical interpretation in the limit $\sigma_1,\sigma_2\rightarrow0$, which is basically the same as that of \eqref{eq:Schwarzschildr} when $\vert\sigma\vert\rightarrow0$. This can be found in \cite{GP}. There it is shown that, after a change of coordinates
\begin{equation}T=\pm r \cosh\theta,\quad R=r\sinh\theta, \quad Z=t,\end{equation}
the metric becomes Minkowski in cylindrical coordinates along the $Z$ axis, namely
\begin{equation}ds^2=-dT^2+dR^2+R^2d\phi^2+dZ^2.\end{equation}
Since $r^2=T^2-R^2$, the hypersurface $r=0$ (which naively represents the region of strong coupling since we have taken $\sigma_1,\sigma_2\rightarrow0$) corresponds to $T=0$, $R=0$ (the worldline of a spacelike particle moving along the $Z$-axis) plus the cylindrical surface $T=\pm R$, which can be understood as a cylindrical wave shrinking to zero size and then expanding again at the speed of light. The resulting configuration may be interpreted as the asymptotic metric, as $r\rightarrow\infty$, of the gravitational field of a tachyon, with the $T=\pm R$ null hypersurfaces corresponding to the horizon $r\approx r_H$. The difference between the solutions \eqref{eq:RNr2} and \eqref{eq:Schwarzschildr} would be, as far as this physical interpretation is concerned, that the tachyon of \eqref{eq:RNr2} carries some charges as dictated by \eqref{Vbh}.

%
%
%

\newpage

\section{Attractor mechanism for topological solutions}

The results of section \ref{sec:TopologicalRN} illustrate the casual structure of the Supergravity class of solutions $C_{3}$, which is in 1-1-1 correspondence with the Supergravity static, spherically symmetric, asymptotically flat black holes of $C_{1}$ and the \emph{hvLif} solutions of $C_{2}/\mathbb{Z}_{2}$. From the very same solution $(U,\phi^{i})\, ,\,\, i=1,\dots,n_{s}$ of the system of differential equations (\ref{eq:e1}), (\ref{eq:Vbh-r0-real}) and (\ref{eq:e3}) we can build three different four-dimensional solutions $s_{1}\in C_{1}$, $s_{2}\in C_{2}/\mathbb{Z}_{2}$ and $s_{3}\in C_{3}$ of the original theory. Since the class $C_{1}$ correspond to spherically symmetric, static, asymptotically flat black holes, the flow of the corresponding scalars may exhibit attractors, or fixed points, at $\tau\rightarrow -\infty$ \cite{Ferrara:1996dd,Ferrara:1997tw,Tripathy:2005qp,Sen:2005wa,Goldstein:2005hq,Bellucci:2007ds,Ferrara:2007qx,Ferrara:2007tu,Ceresole:2007wx,
Ferrara:2008hwa,Bellucci:2008cb,Bellucci:2008jq,Marrani:2010bn}. This is in particular ensured for supersymmetric black holes. Amazingly enough, the scalars of the related solution $s_{3}$ are the very same of the scalars as those of $s_{1}$, so the scalars of $s_{3}$ will have fixed points if and only if the scalars of $s_{1}$ also have them.

The previous considerations prove the attractor mechanism for a subset $C^{\rm Att}_{3} \subset C_{3}$ such that the related solutions in $C_{1}$ also exhibits an attractor mechanism. However, in the case of solutions in $C_{3}$, the scalars are not fixed at an event horizon, since the solution does not have any, but instead they are fixed at the Killing horizon.

\section*{Acknowledgments}

The authors would like to thank P. Bueno, L. J. Garay, R.B. Mann, T. Ort\'in and J. Podolsky for useful discussions and comments.
AdlCD acknowledges financial support from MINECO (Spain) projects numbers FIS2011-23000, FPA2011-27853-C02-01 and Consolider-Ingenio MULTIDARK CSD2009-00064.
AdlCD also acknowledges financial support from Marie Curie - Beatriu de Pin\'os contract BP-B00195, Generalitat de Catalunya and ACGC, University of Cape Town.
MM acknowledges support from the Excellence Campus M. Sc. in Theoretical Physics scholarship.
CSS thanks the Stanford Institute for Theoretical Physics for its hospitality during the earlier stages of this work.  CSS has been supported 
by the JAE-predoc grant JAEPre 2010 00613 and acknowledges partial support 
 by the Spanish Ministry of Science and Education grant FPA2009-07692, the Comunidad de Madrid grant HEPHACOS S2009ESP-1473, and the Spanish Consolider-Ingenio 2010 program CPAN CSD 2007-00042.

\bibliographystyle{JHEP}
\bibliography{References2}
\label{biblio}

\end{document}